\documentclass[pre,amsmath,aps,twocolumn,superscriptaddress]{revtex4-1}
\usepackage{epsfig}
\usepackage{amssymb}

\newcommand{\noi}{\noindent}
\newcommand{\be}{\begin{equation}}
\newcommand{\ee}{\end{equation}}
\newcommand{\bdm}{\begin{displaymath}}
\newcommand{\edm}{\end{displaymath}}

\begin{document}

\title{Unusual Response to a Localized Perturbation in a Generalized Elastic Model}

\author{Alessandro Taloni}
\affiliation{School of Chemistry, Tel Aviv University, Tel Aviv 69978, Israel}
\author{Aleksei Chechkin}
\affiliation{School of Chemistry, Tel Aviv University, Tel Aviv 69978, Israel}
\affiliation{Akhiezer Institute for Theoretical Physics, NSC KIPT,
  Kharkov 61108, Ukraine}
\author{Joseph Klafter}
\affiliation{School of Chemistry, Tel Aviv University, Tel Aviv 69978, Israel}

\begin{abstract}
The generalized elastic model encompasses several physical systems
such as polymers, membranes, single file systems, fluctuating surfaces and rough interfaces.  We consider 
the case of an applied localized potential, namely an external
force  acting only on  a single (tagged) probe, leaving the rest of the system unaffected. We derive the
fractional Langevin equation for the tagged probe, as well as for a generic (untagged) probe, where the force is not directly applied. Within the framework of the
fluctuation-dissipation relations, we discuss the unexpected  physical scenarios arising when the force is constant and time periodic, whether or not the hydrodynamic interactions are included in the model. For short times, in case of the constant force,
we show that the average drift is linear in time for long
range hydrodynamic interactions and behaves ballistically
or exponentially for local hydrodynamic interactions.
Moreover, it can be opposite to the direction of external
disturbance for some values of the model's parameters. When the force is time periodic, the effects are macroscopic: the system splits into two distinct spatial regions whose size is proportional to the value of the applied frequency. These two regions are characterized by different amplitudes and phase shifts in the response dynamics. 
\end{abstract}
\maketitle

%
%
\section{Introduction}
Semiflexible and flexible polymeric chains ~\cite{Doi, Rouse, Zimm, Granek, Farge, Caspi, Amblard}, membranes ~\cite{Granek, Freyssingeas, Helfer, membranes-FLE, Zilman}, moving interfaces ~\cite{Edwards, Joanny, Searson, Krug, Krug_1}, growing surfaces ~\cite{surfaces}, single file systems ~\cite{Lizana} are just few  among the linear elastic systems whose time evolution is ruled by the stochastic equation

\begin{equation}
\frac{\partial}{\partial t}\mathbf{h}\left(\vec{x},t\right)=\int d^dx'\Lambda\left(\vec{x}-\vec{x}'\right)\frac{\partial^z }{\partial\left|\vec{x}'\right|^z }\mathbf{h}(\vec{x}',t)+\boldsymbol\eta\left(\vec{x},t\right),
\label{GEM}
\end{equation}

\noi termed as generalized elastic model  ~\cite{our-PRL,our-PRE}. In its general formulation, the GEM (\ref{GEM}) is given for a $D$-dimensional stochastic field $\mathbf{h}$ defined in the $d$-dimensional infinite space $\vec{x}$.  The Gaussian random noise source
satisfies the fluctuation-dissipation (FD) relation  $\langle\eta_{j}\left(\vec{x},t\right)\eta_{k}\left(\vec{x}',t'\right) \rangle
=  2k_BT\Lambda\left(\vec{x}-\vec{x}'\right)\delta_{j\,k}\delta(t-t')
$ ($j,k\in[1,D]$), where $\Lambda\left(\vec{r}\right)=1/\left|\vec{r}\right|^\alpha$ corresponds to the hydrodynamic friction kernel whose the Fourier transform is
$\Lambda\left(\vec{q}\right)
=\frac{(4\pi)^{d/2}}{2^{\alpha}}\frac{\Gamma\left((d-\alpha)/2\right)}{\Gamma\left(\alpha/2\right)}\left|\vec{q}\right|^{\alpha-d}=A\left|\vec{q}\right|^{\alpha-d}
$,  if $\frac{d-1}{2}<\alpha<d$. The fractional derivative appearing in the right hand side of Eq.(\ref{GEM}) is commonly defined as fractional Laplacian ($\frac{\partial^z}{\partial\left|\vec{x}\right|^z}:=-\left(-\nabla^2\right)^{z/2}$~\cite{Samko}), and is expressed in terms of its Fourier transform ${\cal F}_{\vec{q}}\left\{\frac{\partial^z}{\partial\left|\vec{x}\right|^z}\right\}\equiv-\left|\vec{q}\right|^z$ ~\cite{Zazlawsky}. A specific choice of the numerical values of the  parameters characterizing Eq.(\ref{GEM}), namely $D,d,\alpha$ and $z$, leads to each one of the systems aforementioned (see ~\cite{our-PRE} for a detailed description). However, a main partition between the systems obeying to Eq.(\ref{GEM}) can be done according to whether or not the  hydrodynamic interactions can be considered long range. Indeed, if the hydrodynamics is only local, i.e. $\Lambda\left(\vec{r}\right)\equiv\delta\left(\vec{r}\right)$, it must be formally set $\alpha=d$ and $A=const$ in its Fourier transform expression $\Lambda\left(\vec{q}\right)$.
For instance, setting  $D=1$, $z=4$, $\alpha=1$ and $d=2$ in (\ref{GEM}) corresponds to the stochastic equation for the height of a free fluid membrane floating in a solvent ~\cite{Granek, Freyssingeas, Helfer, membranes-FLE, Zilman}, while for $D=3$, $z=2$, $\Lambda\left(\vec{r}\right)\equiv\delta\left(\vec{r}\right)$ Eq.(\ref{GEM}) governs the monomer's dynamics in a Rouse polymer ~\cite{Rouse}.

\subsection{Fractional Langevin equation scheme}
\noi Tracing the derivation firstly obtained for the single file model ~\cite{Lizana},   recently we have shown ~\cite{our-PRL} how to derive the stochastic equation governing the motion of a \emph{tracer} or \emph{probe particle} placed at the system's position $\vec{x}$, that is the fractional Langevin equation (FLE) 

\be
K^+D_C^{\beta}\mathbf{h}\left(\vec{x},t\right)=\boldsymbol\zeta\left(\vec{x},t\right)
\label{FLE},
\ee

\noi with $\beta=\frac{z-d}{z+\alpha-d}$ and
$K^+=\pi^{d/2-1}\sin\left(\pi\beta\right)\frac{\Gamma(d/2)}{2^{1-d}A^\beta}(z+\alpha-d)$.
Here the fractional Caputo derivative ~\cite{Caputo,Podlubny} is
defined as
$D_C^{\beta}\phi(t)=\frac{1}{\Gamma\left(1-\beta\right)}\int_{-\infty}^tdt'\left(t-t'\right)^{-\beta}\frac{d}{dt'}\phi\left(t'\right)$
$(0<\beta<1)$ and its Fourier transform is given by ${\cal F}_{\omega}\left\{D_C^{\beta}\phi(t)\right\}=(-i\omega)^\beta\phi(\omega)$. The  fractional Gaussian noise
$\boldsymbol\zeta\left(\vec{x},t\right)$ satisfies the \emph{second}
fluctuation-dissipation (FD) relation

\be
\langle \zeta_j\left(\vec{x},t\right)
\zeta_k\left(\vec{x},t'\right)\rangle=k_BT \frac{K^+}{\Gamma\left(1-\beta\right)\left|t-t'\right|^{\beta}}\delta_{j,k}.
\label{FLE-FDT}
\ee

\noi The FLE representation (\ref{FLE}) is a mere change of wording compared to that furnished by (\ref{GEM}).
However, rephrasing the system's dynamics may result to be very helpful in the understanding of the physical picture subtenting the tracer's motion.
This will be amply demonstrated by the following analysis.

\subsection{Generalized elastic model with localized potential}
In this article  we generalize the outlined framework by deriving  the tracer's FLE when a localized potential is applied to the probe in $\vec{x}^{\star }$ (\emph{tagged} probe). Our starting point is the following GEM

\begin{equation}
\begin{array}{l}
\frac{\partial}{\partial t}\mathbf{h}\left(\vec{x},t\right)=\int d^dx'\Lambda\left(\vec{x}-\vec{x}'\right)\times\\
\     \ \left[\frac{\partial^z }{\partial\left|\vec{x}'\right|^z
}\mathbf{h}(\vec{x}',t)+\mathbf{F}\left\{\mathbf{h}(\vec{x}',t),t\right\}\delta(\vec{x}'-\vec{x}^{\star
})\right]+\boldsymbol\eta\left(\vec{x},t\right),
\label{GEM_potential}
\end{array}
\end{equation}

\noi with the local applied force $\mathbf{F}$. \noi For such
systems the tagged probe's FLE  has been derived in two particular
cases: membranes ~\cite{membranes-FLE} ($z=4$, $\alpha=1$, $d=2$),
where the applied harmonic potential was introduced to mimic the
action of an optical/magnetic tweezer; and single file systems
~\cite{Lizana} ($z=2$, $\alpha=d=1$), where three types of forces
were analyzed: constant, time-oscillating and hookean. 

\noi In Sec.\ref{sec:FLE} we  draw the FLE for the \emph{untagged}
tracer ($\vec{x}\neq\vec{x}^{\star }$), besides the tagged one. Our
analysis reveals different surprising regimes attained by the untagged probe in $\vec{x}$, validating the Kubo fluctuation relations  (KFR) in presence of a constant and
time-periodic force in $\vec{x}^{\star }$. Quoting Kubo
~\cite{Kubo}, these relations state general relationships
``between the response of a given system to an external disturbance
and the internal fluctuation of the system in the absence of the
disturbance''. In Sec.\ref{sec:const_force} we study the case of an  applied constant force. Indeed, while the tagged probe response is sublinear in time, the untagged probe's is always different at short times: it is linear for systems characterized by long range hydrodynamic interactions, and it is ballistic (with a sign depending on the model's parameters) or exponential in case of local hydrodynamics. Asymptotically, the tagged and untagged probes undergo the same dynamical behavior. In Sec.\ref{sec:periodic_force} the force is time periodic, in this case the macroscopic effects are persistent in time. Indeed the system splits into two macroregions whose size is defined by the value of the applied frequency $\omega_0$: the  responses of these regions  markedly differs in amplitude and phase, allowing to make experimentally testable predictions on the viscoleastic properties of the system. In Appendix \ref{app:Champeney} we furnish the $d$-dimensional expression for the Fourier transform and its inverse. In Appendix \ref{app:Olver} we report the theorem for the Laplace method for solving asymptotic integrals. In Appendix \ref{app:Erdelyi} the formula for asymptotic solution of Fourier integrals is furnished.

%
%
\section{Fractional Langevin Equation}
\label{sec:FLE}

We start from Eq.(\ref{GEM_potential}) and give its solution in the Fourier space. We first define the Fourier transform in space and time as  $\mathbf{h}\left(\vec{q},\omega\right)=\int_{-\infty}^{+\infty}d^dx \int_{-\infty}^{+\infty}dt\,
\mathbf{h}\left(\vec{x},t\right)\,e^{-i\left(\vec{q}\cdot\vec{x}-\omega t\right)}$, and introduce the notation for the time-Fourier transform of the force: ${\cal F}_{\omega}\left\{\mathbf{F}\left\{\mathbf{h}(\vec{x}^\star,t),t\right\}\right\}\equiv\int_{-\infty}^{+\infty}dt\,\mathbf{F}\left\{\mathbf{h}(\vec{x}^\star,t),t\right\}
\,e^{i\omega t}$. We proceed to the FLE derivation for the untagged tracer in $\vec{x}$. The solution of Eq.(\ref{GEM_potential}) reads

\be
\mathbf{h}\left(\vec{q},\omega\right)=\frac{A\,{\cal F}_{\omega}\left\{\mathbf{F}\left\{\mathbf{h}(\vec{x}^\star,t),t\right\}\right\}e^{-i\vec{q}\cdot\vec{x}^\star}}{\left|\vec{q}\right|^{d-\alpha}\left(-i\omega+A\left|\vec{q}\right|^{\gamma/2}\right)}+\frac{\boldsymbol{\eta}\left(\vec{q},\omega\right)}{-i\omega+A\left|\vec{q}\right|^{\gamma/2}},
\label{sol_FF}
\ee

\noi where we made use of the short notation $\gamma=2(z+\alpha-d)$. After multiplying  both sides of the equation by $K^+(-i\omega)^{\beta}$ and, inverting the Fourier transform in space by means of Eq.(\ref{Champa_antitrasf}), we get

\be
\begin{array}{l}
K^+(-i\omega)^{\beta}\mathbf{h}\left(\vec{x},\omega\right)=\\
\     \ {\cal F}_{\omega}\left\{\mathbf{F}\left\{\mathbf{h}(\vec{x}^\star,t),t\right\}\right\}\Theta\left(\left|\vec{x}-\vec{x}^\star\right|,\omega\right)+\boldsymbol{\zeta}\left(\vec{x},\omega\right),
\label{FLE_x_F}
\end{array}
\ee

\noi where the non-Markovian Gaussian noise  is introduced as ~\cite{our-PRL}

\be
\boldsymbol{\zeta}\left(\vec{x},\omega\right)=\int_{-\infty}^{+\infty}d\vec{x}'\boldsymbol{\eta}\left(\vec{x}',\omega\right)\Phi\left(\left|\vec{x}'-\vec{x}\right|,\omega\right)
\label{sol_FF_2term}.
\ee

\noi The functions $\Theta$ and $\Phi$ appearing in (\ref{FLE_x_F}) and in (\ref{sol_FF_2term}) are expressed as $\Theta\left(\left|\vec{x}\right|,\omega\right)=A{\cal I}_{\alpha-d/2}\left(\left|\vec{x}\right|,\omega\right)$ and $\Phi\left(\left|\vec{x}\right|,\omega\right)={\cal I}_{d/2}\left(\left|\vec{x}\right|,\omega\right)$, where

\be
\begin{array}{l}
{\cal I}_{\lambda}\left(\left|\vec{x}\right|,\omega\right)=\\
\      \ \frac{K^+\left|\vec{x}\right|^{1-d/2}}{(2\pi)^{d/2}}(-i\omega)^{\beta}\int_{0}^{+\infty}d\left|\vec{q}\right|\frac{\left|\vec{q}\right|^{\lambda}J_{d/2-1}\left(\left|\vec{q}\right|\left|\vec{x}\right|\right) }{-i\omega+A\left|\vec{q}\right|^{\gamma/2}}.
\label{Theta_Phi_F}
\end{array}
\ee

\noi Their expression in time yields

\be
\begin{array}{l}
{\cal I}_{\lambda}\left(\left|\vec{x}\right|,t\right)=\frac{K^+\left|\vec{x}\right|^{1-d/2}}{(2\pi)^{d/2}} \,{D_C^{\beta}}\int_{0}^{+\infty}d\left|\vec{q}\right|\left|\vec{q}\right|^{\lambda}\times\\
\              \
J_{d/2-1}\left(\left|\vec{q}\right|\left|\vec{x}\right|\right)e^{-A\left|\vec{q}\right|^{\gamma/2}t}\theta(t),
\label{Theta_Phi}
\end{array}
\ee

\noi replacing $\lambda=\alpha-d/2$ and $\lambda=d/2$ respectively, and where $\theta(t)$ is the Heaviside step function, and $J_{d/2-1}$  denotes the Bessel function of  order $d/2-1$. We point out that both functions $\Theta$ and $\Phi$ coincide in the case of local hydrodynamics ($\alpha=d$ and $A=const$). The physical picture behind the  above mathematical derivation gets clear after inverting in time the equations (\ref{FLE_x_F}) and (\ref{sol_FF_2term}):

\be
\begin{array}{l}
K^+D_C^{\beta}\mathbf{h}\left(\vec{x},t\right)=\\
\     \     \int_{-\infty}^{t}dt'\mathbf{F}\left\{\mathbf{h}(\vec{x}^\star,t'),t'\right\}\Theta\left(\left|\vec{x}-\vec{x}^\star\right|,t-t'\right)+\boldsymbol{\zeta}\left(\vec{x},t\right),
\label{FLE_x}
\end{array}
\ee

\noi with

\be
\boldsymbol{\zeta}\left(\vec{x},t\right)=\int_{-\infty}^{+\infty}d\vec{x}'\int_{-\infty}^{t}dt'\boldsymbol{\eta}\left(\vec{x}',t'\right)\Phi\left(\left|\vec{x}'-\vec{x}\right|,t-t'\right).
\label{sol_2term}
\ee

\noi   Thus $\Theta\left(\left|\vec{x}-\vec{x}^\star\right|,t-t'\right)$ can be seen as the propagator carrying the external perturbation, exerted at the point $\vec{x}^\star$ at time $t'$,  to the point $\vec{x}$ at time $t$. Likewise, the function $\Phi\left(\left|\vec{x}-\vec{x}'\right|,t-t'\right)$ represents the propagator of the Brownian random source $\boldsymbol{\eta}\left(\vec{x},t\right)$ from the point $\vec{x}'$ to the point $\vec{x}$ in the time elapsed between $t'$ and $t$.

\noi We now turn to the derivation of the FLE for the tagged tracer. In this case it will be sufficient take the limit $\vec{x}\to \vec{x}^\star$ in (\ref{FLE_x_F}), which corresponds to set $\left|\vec{x}\right|=0$ in the ${\cal I}_{\alpha-d/2}\left(\left|{\vec x}\right|,\omega\right)$  expression (\ref{Theta_Phi_F}). To this end we recall that the Bessel function expansion for small argument is
$J_{d/2-1}(r)\sim\frac{1}{\Gamma(d/2)}\left(\frac{2}{r}\right)^{1-d/2}$ ~\cite{Abramowitz}), from which follows $\Theta\left(0,\omega\right)\equiv A{\cal I}_{\alpha-d/2}\left(0,\omega\right)=1$ after straightforward passages. Substituting in (\ref{FLE_x_F}) and inverting in time gives the  FLE expression  for the probe particle placed at $\vec{x}^\star$ :

\be
K^+D_C^{\beta}\mathbf{h}\left(\vec{x}^\star,t\right)=\mathbf{F}\left\{\mathbf{h}(\vec{x}^\star,t),t\right\}+\boldsymbol{\zeta}\left(\vec{x}^\star,t\right).
\label{FLE_xstar}
\ee

\noi  In the upcoming sections we analyze the cases of a constant  and a time-periodic external force applied  in $\vec{x}^\star$.

\begin{figure}
\centerline{\includegraphics[width=.3\textwidth]{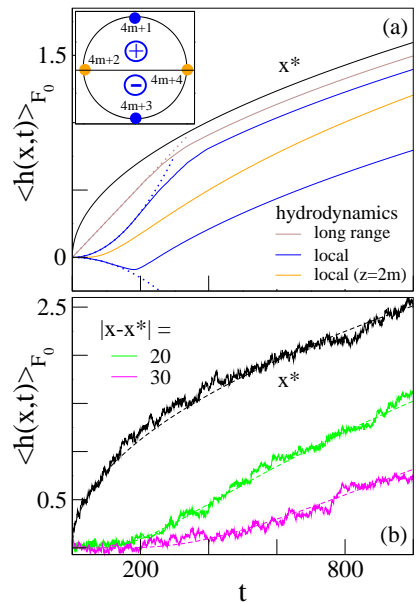}}
\caption{(Color online) Constant force $F_0$. (a) Schematic representation 
of the response (\ref{constant_force_average_drift}). Black (upper) line: 
Einstein relation (\ref{GER}) for the tagged probe. Grey line (second line 
from the top): average drift for the untagged tracer in presence of long 
ranged hydrodynamic interactions, dotted line represents the linear expression 
(\ref{constant_force_drift_S_times_hydro}) at small times. Blue lines 
(third and fifth line from the top): $\langle h(x,t)\rangle_{F_0}$ for local 
hydrodynamics, upper (the third) and bottom (the fifth) curves stand for the 
positive and negative responses 
(\ref{constant_force_drift_S_times_local}) for $t<\tau_{diff}$ and 
$4m<z<4m+2$ and $4m+2<z<4m+4$, respectively (dotted lines);  the middle orange curve 
(fourth from the top) represents the drift for $z=2m$, which is
exponentially small $\propto e^{-t/\tau_{diff}}$ at small times 
(see also panel (b)). 
Inset: representation of the phase $z\pi/2$ in (\ref{constant_force_drift_S_times_local}), 
for z corresponding to the upper plane the drift has the same direction as 
$F_0$ while it is opposite in the bottom; exponentially small drift arises 
at $z=2m$ (orange (horizontal) solid circles). 
(b) Average probe's drift for the Edward-Wilkinson 
chain ($D=1, z=2, \alpha=d=1$, and $A=1/\xi$, where $\xi$ is the damping). 
Simulations are carried out submitting the tagged tracer to a force $F_0=0.33$ 
and detecting the average drift for different probes (solid green (middle) and magenta (bottom) curves). 
Dashed lines represent the theoretical expressions (\ref{constant_force_average_drift}): 
$\langle h\left(x,t\right)\rangle_{F_0}=F_0\left[\sqrt{t/(\pi\xi)}e^{-y^2}-
\left(\left|x-x^\star\right|/2\right)\,erfc\left(y\right)\right]$ 
with $y=\left|x-x^\star\right|\sqrt{\xi/(4t)}$.   
Statistical averages were taken over 2000 realizations. 
Other simulation's parameters are $\xi=1.0$ and $k_BT=1.0$. The expression of the 
diffusion time $\tau$, that entails the transition from short  to long time behavior, 
is given by $\tau=\left|x-x^\star\right|^2\xi$, which is $\tau=400$ for the green (middle) curve and $\tau=900$ for the magenta (bottom) curve.}
\label{fig.1}
\end{figure}

\section{Constant force}
\label{sec:const_force}

Let $F_0$ represents the force along one
direction only, say $F_0\equiv F_j$:
$\mathbf{F}\left\{\mathbf{h}(\vec{x},t),t\right\}=F_0\theta(t)$. We are interested in the average drifts  $\langle h\left(\vec{x},t\right)\rangle_{F_0}$ and  $\langle h\left(\vec{x}^\star,t\right)\rangle_{F_0}$ where we  dropped the index $j$. By averaging both Eq.(\ref{FLE_x}) and
(\ref{FLE_xstar}) and plugging in the definition  of $\Theta\left(\left|\vec{x}-\vec{x}^\star\right|,t-t'\right)$ one has

\be
\begin{array}{l}
\langle h\left(\vec{x},t\right)\rangle_{F_0}=\frac{A\left|\vec{x}-\vec{x}^\star\right|^{1-d/2}}{(2\pi)^{d/2}}F_0\int_0^t dt'\times\\
\           \ \int_{0}^{+\infty}d\left|\vec{q}\right|\left|\vec{q}\right|^{\alpha-d/2}J_{d/2-1}\left(\left|\vec{q}\right|\left|\vec{x}-\vec{x}^\star\right|\right)e^{-A\left|\vec{q}\right|^{\gamma/2} t'},\\
\langle h\left(\vec{x}^\star,t\right)\rangle_{F_0}=\frac{F_0}{K^+\Gamma\left(1+\beta\right)}\,t^{\beta}
.
\label{constant_force_average_drift}
\end{array}
\ee

\noi Comparing the tracers' responses in (\ref{constant_force_average_drift}) it is clear that  the Einstein relation only holds for the tagged probe, i.e.

\be
\langle h\left(\vec{x}^{\star},t\right)\rangle_{F_0}=\frac{ \langle\delta^2h(t)\rangle}{2k_BT}F_0.
\label{GER}
\ee

\noi As a matter of fact, we recall that  the mean square displacement in the absence of external force is given by $\langle\delta^2 h(t)\rangle=2\frac{k_BT}{K^+\Gamma(1+\beta)}t^{\beta}$ ~\cite{our-PRL}. On the other hand the untagged tracer fulfils the more general KFR ~\cite{UMB, Villamaina},  
and its time behavior presents very interesting features.
Indeed, the analysis of  Eq.(\ref{constant_force_average_drift}) leads us to the following conclusions: (a)  the drift $\langle h\left(\vec{x},t\right)\rangle_{F_0}$ attains two different behaviors for times larger and smaller than the correlation time $\tau=\left|\vec{x}-\vec{x}^\star\right|^{\gamma/2}/A$; (b) for $t< \tau$, the response is dissimilar whether the hydrodynamic interactions are considered to be long range or local.

\begin{itemize}
\item $\mathbf{t<} \boldsymbol{\tau}$. \emph{Long range hydrodynamic interactions.--} In this case the integral over $\vec{q}$ appearing in (\ref{constant_force_average_drift}) can be performed asymptotically, the solution to the main order is

\be
\langle h\left(\vec{x},t\right)\rangle_{F_0}\sim\frac{\Gamma(\alpha/2)}{\Gamma((d-\alpha)/2)}\frac{2^{\alpha-d}}{\pi^{d/2}}
\frac{A}{\left|\vec{x}-\vec{x}^\star\right|^\alpha} F_0t.
\label{constant_force_drift_S_times_hydro}
\ee

\noi \emph{Local hydrodynamic interactions.--} We put $\alpha=d$ (and hence $\gamma=2z$) in Eq.(\ref{constant_force_average_drift}). The  integral over $\vec{q}$ is  evaluated in Appendix \ref{app:Erdelyi}. The untagged probe's drift expression  in Eq.(\ref{constant_force_average_drift}) is given by

\be
\begin{array}{l}
\langle h\left(\vec{x},t\right)\rangle_{F_0}\sim\\
\		\ \frac{2^{z-2}}{\pi^{1+d/2}}z\Gamma(\frac{z}{2})\Gamma(\frac{z+d}{2})\sin\left(\frac{z\pi}{2}\right)\frac{A}{\left|\vec{x}-\vec{x}^\star\right|^{d}} \frac{F_0t^2}{\tau},
\label{constant_force_drift_S_times_local}
\end{array}
\ee

\noi for $z\neq 2m$, with $m\in \mathbb{N}$. The results
(\ref{constant_force_drift_S_times_hydro}) and
(\ref{constant_force_drift_S_times_local}) suggest that the untagged
probe moves in average as a free Brownian and ballistic particle
respectively, under the influence of an external effective force whose
amplitude is inversely proportional to the distance
$\left|\vec{x}-\vec{x}^\star\right|$. However, the ballistic picture
is reductive in the case of local hydrodynamics. As a matter of fact
for $2+4m<z<4+4m$ the response of the probe is opposite to the
external disturbance $F_0$, while for  $4m<z<2+4m$ they have the
same sign. For $z=2m$ the response is slower than any power so that
we expect $\langle h\left(\vec{x},t\right)\rangle_{F_0}\propto F_0
t^{\beta+1}/\tau \,e^{-\tau/t}$ (up to numerical
prefactor in the exponential function). These surprising and
counterintuitive results are summarized in Figure \ref{fig.1}(a).

\noi To take an example, let us discuss the situation of growing surfaces. The cases $z=2$, $z=3$ and $z=4$ ($d=1$) refer to different types of atomic diffusion on a crystalline surface ~\cite{surfaces}. The response of the step $h\left(x,t\right)$ (the line boundary at which the surface changes height) to $F_0$ grows in time exponentially in the case of attachment-detachment diffusion ($z=2$):  in this instance the complete solution of the first of  Eq.(\ref{constant_force_average_drift}) is achieved, yielding $\langle h\left(x,t\right)\rangle_{F_0}=F_0\left[\sqrt{t/(\pi\xi)}e^{-y^2}-\left(\left|x-x^\star\right|/2\right)\,erfc\left(y\right)\right]$ (Figure \ref{fig.1}(b)), where $y=\left|x-x^\star\right|\sqrt{\xi/(4t)}$, $A=1/\xi$ and $\xi$ is the viscous coefficient; its behavior at small times is found to be $\langle h\left(x,t\right)\rangle_{F_0}\sim F_0(t/\xi)^{3/2}\left|x-x^\star\right|^{-2} e^{-\xi\left|x-x^\star\right|^2/(4t)}$ . For the terrace diffusion ($z=3$) Eq.(\ref{constant_force_drift_S_times_local}) gives a negative drift, i.e $\langle h\left(x,t\right)\rangle_{F_0}\sim-\frac{6}{\pi}\frac{F_0(At)^2}{\left|x-x^\star\right|^{4}}$. Finally if $z=4$, the so called periphery diffusion, one recovers the exponential growth at short times, i.e $\propto F_0 t^{7/4}\left|x-x^\star\right|^{-4}e^{-\left|x-x^\star\right|^{4}/(At)}$.

\item $\mathbf{t>} \boldsymbol{\tau}$. The integral in $\left|\vec{q}\right|$ appearing in the first of  Eqs.(\ref{constant_force_average_drift}) can be solved by using the Laplace method (see  Appendix(\ref{app:Olver})): $\langle h\left(\vec{x},t\right)\rangle_{F_0}=\frac{F_0}{K^+\Gamma\left(1+\beta\right)}\,t^{\beta}$. Thus the Einstein relation (\ref{GER}) is  then regained only when $t>\tau$, namely when the correlation lenght $\xi=(At)^{2/\gamma}$ exceeds the distance $\left|\vec{x}-\vec{x}^\star\right|$. The transient violation of Einstein relation contrasts with the second FD Eq.(\ref{FLE-FDT}), which is always fulfilled: the larger the distance $\left|\vec{x}-\vec{x}^\star\right|$, the longer the transient.
\end{itemize}

We can summarize the results obtained in this section in the following compact form. The average drift is cast as

\be
\langle h\left(x,t\right)\rangle_{F_0} =F_0\left|\vec{x}-\vec{x}^\star\right|^{z-d}f\left[\frac{t}{\tau}\right]
\label{scaling_average_drift}
\ee

\noi The scaling function $f\left[u\right]$ exhibits two distinct behaviours whether $u\ll 1$ or $u\gg 1$. From (\ref{constant_force_drift_S_times_hydro}) and (\ref{constant_force_drift_S_times_local}), it turns out that when $u\ll 1$
 
\be
f\left[u\right]
\begin{array}{ccc}
\sim\frac{2^{\alpha-d/2}}{\pi^{d/2}}\frac{\Gamma\left(\frac{\alpha}{2}\right)}{\Gamma\left(\frac{d-\alpha}{2}\right)}u &   & i)\\
\sim\frac{2^{z-2}}{\pi^{1+d/2}}z\sin\left(\frac{z\pi}{2}\right)\Gamma\left(\frac{z}{2}\right)\Gamma\left(\frac{z+d}{2}\right)u^2 &   & ii)\\
\propto u^{\beta+1}e^{-\frac{1}{u}} &   & iii)
\end{array}
\label{f_sht_th}
\ee

\noi for $i)$ long range, $ii)$ local ($z\neq 2m$) and $iii)$ local ($z= 2m$) hydrodynamic interactions, respectively. When $u\gg 1$ we have invariably 

\be
f\left[u\right]\simeq\frac{1}{2^{d-1}\pi^{d/2}}\frac{\Gamma(1-\beta)}{(z-d)\Gamma\left(\frac{d}{2}\right)}u^\beta
\label{f_lot_th}.
\ee

\noi Furthermore, it can be shown ~\cite{EPL} that the following relation holds rigorously for both tagged and untagged tracers

\be
\langle h\left(\vec{x},t\right)\rangle_{F_0}=\frac{ \langle\delta h(\vec{x},t)\delta h(\vec{x}^\star,t)\rangle}{2k_BT}F_0,
\label{GER_new}
\ee

\noi which, in turns, encompasses the Einstein relation (\ref{GER}) and its generalization, namely the KFR.

\section{Time periodic force}
\label{sec:periodic_force}

We now consider the force $ \mathbf{F}\left\{\mathbf{h}(\vec{x},t),t\right\}=F_0\cos(\omega_0t)$ and from (\ref{FLE_x}) and (\ref{FLE_xstar}) we have

\be
\begin{array}{l}
K^+D_C^{\beta}h\left(\vec{x},t\right)=
F_0\Re e\left[\int_{-\infty}^{+\infty}dt'e^{-i\omega_0t'}\Theta\left(\left|\vec{x}-\vec{x}^\star\right|,t-t'\right)\right]+\\
\     \
\zeta\left(\vec{x},t\right),\\
K^+D_C^{\beta}h\left(\vec{x}^\star,t\right)=
F_0\,\Re e\left[e^{-i\omega_0t}\right]+\zeta\left(\vec{x}^\star,t\right).
\label{FLE_force_periodic}
\end{array}
\ee

\noi In this case we  study the complex mobilities or admittances $\mu\left(\vec{x},\omega_0\right)$ and $\mu\left(\vec{x}^\star,\omega_0\right)$ which are defined through  the linear response  relations ~\cite{Kubo}

\be
\begin{array}{l}
\langle v\left(\vec{x},t\right)\rangle_{F_0}=\Re e\left[ \mu\left(\vec{x},\omega_0\right)F_0e^{-i\omega_0 t}\right],\\
\langle v\left(\vec{x}^\star,t\right)\rangle_{F_0}=\Re e\left[\mu\left(\vec{x}^\star,\omega_0\right)F_0
e^{-i\omega_0t}\right].
\label{periodic_force_v_linear}
\end{array}
\ee

\noi Both tracers fulfill the generalized Green-Kubo relation 

\be
\langle v\left(\vec{x},\omega_0\right)
v\left(\vec{x}^\star,\omega_0'\right)\rangle=4\pi\delta(\omega_0+\omega_0')k_BT\Re e \left[\mu\left(\vec{x},\omega_0\right)\right],
\label{1-FDT}
\ee

\noi where the complex mobilities follow from Eq.(\ref{FLE_force_periodic}) through the definition (\ref{periodic_force_v_linear}),

\be
\begin{array}{l}
\mu\left(\vec{x},\omega_0\right)=\frac{A\left|\vec{x}-\vec{x}^\star\right|^{1-d/2}}{(2\pi)^{d/2}}\int_{0}^{+\infty}d\left|\vec{q}\right|\left|\vec{q}\right|^{\alpha-d/2}\times\\
\         \
J_{d/2-1}\left(\left|\vec{q}\right|\left|\vec{x}-\vec{x}^\star\right|\right)\frac{-i\omega_0}{-i\omega_0+A\left|\vec{q}\right|^{\gamma/2}},\\
\mu\left(\vec{x}^\star,\omega_0\right)=
\frac{\omega_0^{1-\beta}}{K^+}\, e^{-i(1-\beta)\frac{\pi}{2}},
\label{periodic_force_mobilities}
\end{array}
\ee

\noi and the unperturbed velocity correlation function on the left hand side of Eq.(\ref{1-FDT}) is obtained from Eq.(\ref{FLE}). In particular, we can write the velocity autocorrelation function in the frequency domain as $\langle v\left(\vec{x}^\star,\omega\right)
v\left(\vec{x}^\star,\omega'\right)\rangle =k_BT\frac{\sin(\pi\beta/2)}{K^+}\, \omega^{1-\beta}4\pi\delta(\omega+\omega')$,
and verify the relation (\ref{1-FDT}) for the tagged probe: this is the standard (canonical) formulation of the first FD relation ~\cite{Kubo}.

\noi We now  analyze the low and high frequency behaviors of $\mu\left(\vec{x},\omega_0\right)=\left|\mu\left(\vec{x},\omega_0\right)\right|e^{-i\varphi\left(\vec{x},\omega_0\right)}$.

\begin{itemize}
\item $\boldsymbol{\omega_0\tau<1}$. Changing variable ($y=(A/\omega_0)^{2/\gamma}\left|\vec{q}\right|$) in  $\Re e\left[\mu\left(\vec{x},\omega_0\right)\right]$ and $\Im m\left[\mu\left(\vec{x},\omega_0\right)\right]$ and using the Bessels function's expansion for small arguments, one has $\mu\left(\vec{x},\omega_0\right)=\omega_0^{1-\beta}/K^+\, e^{-i(1-\beta)\frac{\pi}{2}}$.

\item $\boldsymbol{\omega_0\tau>1}$. \emph{Long range hydrodynamic interactions.--} Performing the integral in $\vec{q}$ for  $\Re e\left[\mu\left(\vec{x},\omega_0\right)\right]$ and $\Im m\left[\mu\left(\vec{x},\omega_0\right)\right]$ we obtain for the response's amplitude

\be
\left|\mu\left(\vec{x},\omega_0\right)\right| \sim
\frac{\Gamma(\alpha/2)}{\Gamma((d-\alpha)/2)}\frac{2^{\alpha-d}}{\pi^{d/2}}
\frac{A}{\left|\vec{x}-\vec{x}^\star\right|^\alpha}, \label{periodic_force_mobility_S_freq_hydro}
\ee

\noi while the phase is negligible, i.e $\tan\varphi\left(\vec{x},\omega_0\right)\sim \frac{1}{\omega_0\tau}$. 

\noi \emph{Local  hydrodynamic interactions.--} We employ the same technique as for expression (\ref{constant_force_drift_S_times_local}) achieving, for $z\neq 2m$ ($m\in \mathbb{N}$),

\be
\begin{array}{l}
\left|\mu\left(\vec{x},\omega_0\right)\right| \sim\\
\frac{2^{z-1}}{\pi^{1+d/2}}z\Gamma(\frac{z}{2})\Gamma(\frac{z+d}{2})\left|\sin\left(\frac{z\pi}{2}\right)\right|\frac{A}{\left|\vec{x}-\vec{x}^\star\right|^{d}} (\omega_0\tau)^{-1}
\label{periodic_force_mobility_S_freq_local},
\end{array}
\ee

\noi and $\tan\varphi\left(\vec{x},\omega_0\right)\sim -\omega_0\tau/\cos\left(\frac{z\pi}{2}\right)$. When  $z=2m$, $\left|\mu\left(\vec{x},\omega_0\right)\right|$ is exponentially small and $\varphi\left(\vec{x},\omega_0\right)\simeq(1-\beta)\frac{\pi}{2}-const(\omega_0\tau)^{1/z}$: for instance the exact solution   for $z=2, d=1$ reads $\mu\left(\vec{x},\omega_0\right)=\frac{\sqrt{\omega_0A}}{2}e^{-\sqrt{\frac{\omega_0}{2A}}\left|\vec{x}-\vec{x}^\star\right|}e^{-i\left(\frac{\pi}{4}-\sqrt{\frac{\omega_0}{2A}}\left|\vec{x}-\vec{x}^\star\right|\right)}$.
\end{itemize}

\begin{figure}
\centerline{\includegraphics[width=.4\textwidth]{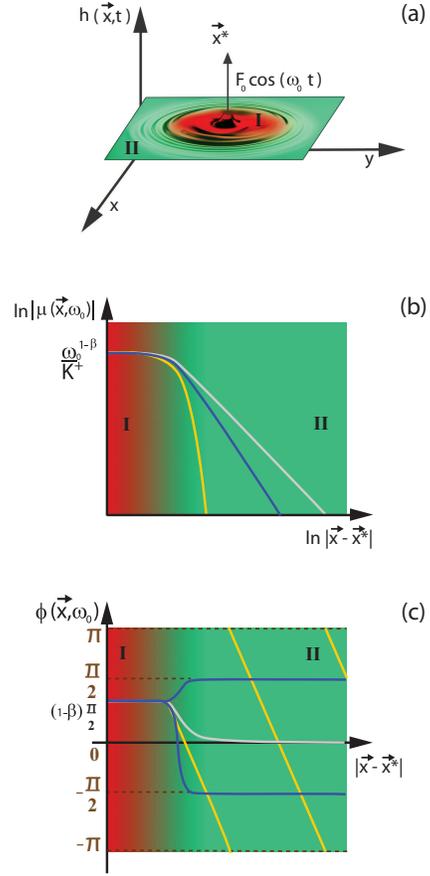}}
\caption{(Color online) Time-periodic force $F_0\cos(\omega_0t)$. (a) 3D rendering of  a membrane described by (\ref{GEM}), under the effect of an applied time-periodic force in $\vec{x}^\star$ (black arrow). $h\left(\vec{x},t\right)$ represents the height of the fluctuating membrane on a 2-dimensional substrate ($\vec{x}=(x,y)$). Regions I and II correspond respectively to the inner  and outer region  in which the membrane separates when the force is applied. The color code, red for region I and green for region II, has been drawn for the reader's convenience: increasing the frequency $\omega_0$ entails the shrinkage of the red region (I). (b) Schematic representation of the untagged response amplitude $\left|\mu\left(\vec{x},\omega_0\right)\right|$ as a function of the distance $\left|\vec{x}-\vec{x}^\star\right|$: since is a schematic drawing  no scale is needed on the $x$-axis. In region I the universal behavior $\omega_0^{1-\beta}/K^+$ holds for any kind of hydrodynamic interactions. In region II the decay of the response's amplitude is $\propto \left|\vec{x}-\vec{x}^\star\right|^{-\alpha}$ for long range hydrodynamic systems (Eq.(\ref{periodic_force_mobility_S_freq_hydro}), grey (upper) solid line), $\propto \left|\vec{x}-\vec{x}^\star\right|^{-z-d}$ for local hydrodynamic systems with $z\neq 2m$ with $m\in\mathbb{N}$ (Eq.(\ref{periodic_force_mobility_S_freq_local}), blue (middle) solid line), and exponentially fast for local hydrodynamic systems with $z=2m$ (orange (bottom) line). (c) Schematic representation of the untagged response phase $\varphi\left(\vec{x},\omega_0\right)$ as a function of the distance $\left|\vec{x}-\vec{x}^\star\right|$. No scale is needed on the $x$-axis. Region I: the system displays an universal phase delay for long range and local hydrodynamic interactions, i.e. $\varphi\left(\vec{x},\omega_0\right)=(1-\beta)\frac{\pi}{2}$. Region II: For long range hydrodynamic systems the phase is absent, i.e. $\varphi\left(\vec{x},\omega_0\right)\simeq 0$ (grey (middle) solid line); for local hydrodynamics the phase is approximately $\pi/2$ if $1+4m<z<3+4m$ with $m\in\mathbb{N}$ (upper blue line), while it is approximately $-\pi/2$ if $4m<z<1+4m$ or $3+4m<z<4+4m$ (bottom blue line); if the hydrodynamic interactions are local and $z=2m$ the phase shows a linear dependence on the distance $\left|\vec{x}-\vec{x}^\star\right|$, i.e. $\varphi\left(\vec{x},\omega_0\right)\sim -\left|\vec{x}-\vec{x}^\star\right|$
(orange (bright linear) solid line).}
\label{fig.2}
\end{figure}

\noi Let us now analyze in detail the physical scenario emerging from the former analysis. The graphical rendering of the following discussion is presented in Fig.\ref{fig.2}. For a given frequency $\omega_0$ the system is divided into two spatial regions, $\left|\vec{x}-\vec{x}^\star\right|<\left(\frac{A}{\omega_0}\right)^{2/\gamma}$ (I) and $\left|\vec{x}-\vec{x}^\star\right|>\left(\frac{A}{\omega_0}\right)^{2/\gamma}$ (II), characterized by very distinct dynamical phases. In case of long range hydrodynamics, the response of the system's portion closer to the tagged probe (I) shows a dependence of the amplitude $\propto \omega_0^{1-\beta}$ and a phase shift $(1-\beta)\pi/2$ with respect to the applied oscillatory force (already noticed in single-file systems ~\cite{Lizana}). On the other hand, in the  outer region (II) the response's amplitude decays as $\left|\vec{x}-\vec{x}^\star\right|^{-\alpha}$, but almost no phase delay is displayed if compared  to the external force. When the  hydrodynamic interactions are local, although region (I) exhibits the same behaviour as in long range interacting systems, in region (II) the amplitude of the response is smaller and decays faster, namely $\propto\left|\vec{x}-\vec{x}^\star\right|^{-z-d} \omega_0^{-1}$ if $z\neq 2m$  and $\propto e^{-\left|\vec{x}-\vec{x}^\star\right|\omega_0^{1/z}}$ for $z=2m$; the phase shift instead is $\varphi\left(\vec{x},\omega_0\right)\approx\pm\frac{\pi}{2}$ for $z\neq 2m$, and it grows like $-\omega_0^{1/z}\left|\vec{x}-\vec{x}^\star\right|$ if $z=2m$.

%
%
\section{Conclusions}

In this paper we derived the  FLE
representation of the tagged ($\vec{x}^\star$) and untagged
($\vec{x}$) probes' dynamics when a localized potential acts on
$\vec{x}^\star$. We demonstrated the validity of the KFR and generalized Green-Kubo relation
for both tracers, and the ensuing non-trivial physical regimes. This findings have important
experimental and theoretical consequences. 

From the experimental point of view,  the response to
a constant force exerted on a position $\vec{x}^\star$ (implemented by an atomic force microscope by instance) can be detected experimentally within the domain of single-particle tracking ~\cite{Saxton}. Indeed, the motion of the untagged tracer ($\vec{x}$), be an optical label, such as a gold or polystyrene bead, or a fluorescent tag, may provide a direct probe of the viscoelastic properties of the system under study, as well as of its underlining elastic energy. The single-particle possible responses are well schematized in Figure \ref{fig.1}, where the different time behaviors undergone by the average drift are displayed. In particular, we notice the surprising effect for which the untagged tracer moves  opposite to the external force for short times.

\noi On the other hand, our analysis provides a quantitive  clear-cut description of the macroscopic observable effects that a localized perturbation produces on an elastic system modeled by (\ref{GEM}). As a matter of fact, a local  oscillating field separates the systems in two regions, whose size can be tuned by tuning the amplitude of the characteristic frequency $\omega_0$. Moreover, we demonstrated that the  behavior attained by these macroscopic domains is characterized by very different amplitudes and phases, according to the type of interaction and the values of the parameters which set the model (\ref{GEM}). Figure \ref{fig.2}  shows the measurable and testable predictions that our analysis enucleated. By instance, the readout of the effects of  a local perturbation exerted by AFM, could be done using differential confocal microscopy to image the membrane ripples ~\cite{ripple}.

\noi  We believe that these findings have important direct applications in the biosensors design and single-molecule manipulations.

From the theoretical point of view  we have shown that  the FLEs (\ref{FLE_x}) and (\ref{FLE_xstar}) constitutes a powerful and comprehensive dynamical representation of the motion of both tagged and untagged probes. Indeed, this stochastic equation, lying in the class of generalized
Langevin equations ~\cite{Kubo}, can be seen as a stochastic
representation of the general Kubo fluctuation relations  with the single-probe random force
$\zeta(\vec{x},t)$ satisfying the second  FD relation. 

%
%
\appendix

\section{Fourier transform of a d-dimensional isotropic function}
\label{app:Champeney}

The $d$-dimensional Fourier transform of $\phi(\vec{r})$ which is function only of its modulus $\left| \vec{r}\right|$, i.e. $\phi(\vec{r})\equiv \phi(\left| \vec{r}\right|)$, is ~\cite{Champeney}

\be
\begin{array}{l}
\int_{-\infty}^{+\infty}d\vec{r} e^{-i\vec{q}\cdot\vec{r}}\phi(\left| \vec{r}\right|)=\\
(2\pi)^{d/2}\left| \vec{q}\right|^{1-d/2}\int_{0}^{+\infty}d\left| \vec{r}\right| \left| \vec{r}\right|^{d/2}J_{d/2-1}(\left| \vec{q}\right|\left| \vec{r}\right|)\phi(\left| \vec{r}\right|).
\end{array}
\label{Champa_trasf}
\ee

\noi Conversely, its inverse Fourier transform $\phi(\left|\vec{q}\right|)$ is given by

\be
\begin{array}{l}
\int_{-\infty}^{+\infty} \frac{d\vec{q}}{(2\pi)^d} e^{i\vec{q}\cdot\vec{r}}\phi(\left| \vec{q}\right|)=\\
\frac{\left|\vec{r}\right|^{1-d/2}}{(2\pi)^{d/2}}\int_{0}^{+\infty}d\left| \vec{q}\right| \left| \vec{q}\right|^{d/2}J_{d/2-1}(\left| \vec{q}\right|\left| \vec{r}\right|)\phi(\left| \vec{q}\right|).
\end{array}
\label{Champa_antitrasf}
\ee

%
%
\section{Laplace method for asymptotic integrals} 
\label{app:Olver}

We hereby report theorem for the asymptotic solution of  exponential integrals through Laplace method ~\cite{Olver}.

 Consider the  integral

\be
I(t)=\int_a^b dq\, x(q)\,  e^{-p(q)t}.
\label{app:Olver_int}
\ee

\noi If the following hypothesis are fulfilled

$i)$ $p(q)>p(a)$ for any $q\in(a,b)$ and the minimum of $p(q)$ is appraoched only at $a$;

$ii)$ $\frac{dp(q)}{dq}$ and $x(q)$ continuous functions in a neighborrod of $a$, except, possibly, at $a$;

$iii)$ as $q\to a^+$,  $(p(q)-p(a))\sim P\, (q-a)^{\mu}$ and  $x(q)\sim Q\, (q-a)^{\lambda-1}$, where $P, \mu$ and $\lambda$ are positive constant and $Q\in\mathbb{R}$ or $\in\mathbb{C}$;

$iv)$  $I(t)$ is absolutely convergent throughout its range for all sufficiently large  $t$;

\noi then the integral $I(t)$ is

\be
I(t)\simeq\frac{Q}{\mu}\Gamma\left(\frac{\lambda}{\mu}\right)\frac{e^{-p(a)t}}{\left(Pt\right)^{\lambda/\mu}}
\label{app:Olver_int_sol}
\ee

\noi in the limit $t\to\infty$.

\label{Olver_theorem}

\noi The integral in (\ref{constant_force_average_drift}) satisfies the previous hypothesis with $a=0$, $b=\infty$, $P=A$, $Q=\frac{2^{1-d/2}}{\Gamma(d/2)\left|\vec{x}-\vec{x}^\star\right|^{1-d/2}}$, $\mu=\gamma/2$ and $\lambda=\alpha$.

%
%
\section{Asymptotic solution of Fourier integrals} 
\label{app:Erdelyi}

We hereby show how to derive Eq.(\ref{constant_force_drift_S_times_local}) from the corresponding general formulation of the untagged drift in  Eq.(\ref{constant_force_average_drift}). We first apply the  change of variable $y=(At)^{2/\gamma}\left|\vec{q}\right|$, achieving

\be
\begin{array}{l}
\langle h\left(\vec{x},t\right)\rangle_{F_0}=\frac{A\left|\vec{x}-\vec{x}^\star\right|^{1-d/2}}{(2\pi)^{d/2}}F_0\int_0^t dt'\times\\
\         \ \left(\frac{1}{At'}\right)^{\frac{2+d}{\gamma}}\int_{0}^{+\infty}dy \, y^{d/2}J_{d/2-1}\left(\lambda y\right)e^{-y^{z}},
\label{constant_force_average_drift_y}
\end{array}
\ee

\noi where $\lambda = \left|\vec{x}-\vec{x}^\star\right|/(At')^{2/\gamma}$ is a large parameter. The integral over $y$ can be evaluated by expanding the exponential for small arguments, i.e. $e^{-y^{z}}\simeq 1-y^z$: the first term gives zero contribution, while the second is ~\cite{Abramowitz}    

\be
\begin{array}{l}
-\int_{0}^{+\infty}dy \, y^{z+d/2}J_{d/2-1}\left(\lambda y\right)=\\
\         \ \frac{2^{z+\frac{d}{2}-1}}{\pi\lambda^{z+\frac{d}{2}+1}}z\sin\left(\frac{z\pi}{2}\right)\Gamma\left(\frac{z}{2}\right)\Gamma\left(\frac{z+d}{2}\right).
\label{app:inte}
\end{array}
\ee

\noi Alternatively, we can use the value of the improper integral

\be
\int_{0}^{+\infty}dy\, y^{\nu}e^{-i\lambda y} = \frac{\Gamma(\nu+1)}{\lambda^{\nu+1}}e^{-i\frac{\pi }{2}(\nu+1)},
\label{Fourier_int}
\ee

\noi with $\nu>-1$, which can be obtained by the method of summation of improper integrals ~\cite{Hardy}.
Recall that the Bessel function for $d=1$ is $J_{-1/2}\left(x\right)=\sqrt{\frac{2}{\pi x}}\cos x$, for $d=3$ is $J_{1/2}\left(x\right)=\sqrt{\frac{2}{\pi x}}\sin x$ and for $d=2$ $J_{0}\left(x\right)\sim\sqrt{\frac{2}{\pi x}}\cos\left( x-\frac{\pi}{4}\right)$ for large $x$ ~\cite{Abramowitz}. 

\noi The real and imaginary part of the mobility $\mu\left(\vec{x},\omega_0\right)$ (\ref{periodic_force_mobilities}) are obtained in the same way.

%
%
\begin{acknowledgments}

A.C. and J.K. acknowledge the support of Marie Curie IIF programme,
grant ``LeFrac''.
A.T. is deeply indebted with Tod  Bertuzzi (TOY), Gianluca Schneider-Faberi aka ``ercapoccetta'' and Andreas Gebardth for their great work on Figure \ref{fig.2}. A.T. aknowlegdes Dario Villamaina, Andrea Puglisi, Angelo Vulpiani and the TNT group for illuminating discussions.

\end{acknowledgments}
%
%


\begin{thebibliography}{10}

\bibitem{Doi}
Doi M and Edwards SF, {\it The Theory of Polymer  Dynamics } (Clarendon, Oxford, 1986).

\bibitem{Rouse}
Rouse PE,  {\em J. Chem. Phys.} {\bf 21} 1272 (1953).

\bibitem{Zimm}
Zimm BH,  {\em J. Chem. Phys.} {\bf 24} 269 (1956).

\bibitem{Granek}
Granek R, {\em J. Phys. II France} {\bf 7} 1761 (1997).

\bibitem{Farge}
Farge E, Maggs AC,{\em Macromol.} {\bf 26} 5041 (1993).

\bibitem{Caspi}
Caspi A, Elbaum M, Granek R, Lachish A, Zbaida D, {\em Phys. Rev. Lett.} {\bf  80} 1106 (1998).

\bibitem{Amblard}
Amblard F, Maggs AC, Yurke B, Pargellis AN, Leibler S,{\em Phys. Rev. Lett.} {\bf  77} 4470 (1996).


\bibitem{Freyssingeas}
Freyssingeas E, Roux D, Nallet F, {\em J. Phys. II France} {\bf   7} 913 (1997).

\bibitem{Helfer}
Helfer E, Harlepp S, Bourdieu L, Robert J, MacKintosh FC, Chatenay D, {\em Phys. Rev. Lett.} {\bf  85} 457 (2000).

\bibitem{membranes-FLE}
Granek R, Klafter J,{\em Europhys. Lett.} {\bf 56} 15 (2001).

\bibitem{Zilman}
Zilman AG, Granek R, {\em Chem. Phys.} {\bf  284} 195 (2002).


\bibitem{Edwards}
Edwards SF, Wilkinson DR, {\em Proc. R. Soc. London A} {\bf  381} 17 (1982) .

\bibitem{Joanny}
Joanny JF, de Gennes PG, {\em J. Chem. Phys.} {\bf 81} 457 (1984).


\bibitem{Searson}
Searson Rong Li PC, Sieradzki K, {\em Phys. Rev. Lett.} {\bf 74} 1395 (1995).

\bibitem{Krug}
Krug J {\em Scale Invariance, Interfaces and Non-Equilibrium Dynamics}  
  (Plenum, New York, 1995).
  
\bibitem{Krug_1}
Krug J, {\em Adv. Phys.} {\bf 46} 139 (1997).


\bibitem{surfaces}
Toroczkai Z, Williams ED, {\em Phys. Today 52, No.} {\bf 12} 24 (1998).



\bibitem{Lizana}
Lizana L, Ambj\"ornsson T, Taloni A,  Barkai E, Lomholt M, {\em Phys. Rev. E} {\bf 81} 051118 (2010).


\bibitem{our-PRL}
Taloni A, Chechkin A, Klafter J, {\em Phys. Rev. Lett.} {\bf 104} 160602 (2010).

\bibitem{our-PRE}
Taloni A, Chechkin A, Klafter J, {\em Phys. Rev. E} {\bf  82} 061104 (2010).

\bibitem{Samko}
Samko SG, Kilbas AA, Marichev OI, \emph {Fractional Integrals
  and Derivatives, Theory and Applications} (Gordon and Breach,
  Amsterdam, 1993).

\bibitem{Zazlawsky}
Saichev A, Zaslavsky G, {\em Chaos} {\bf  7} 753 (1997).

\bibitem{Caputo} Caputo M, {\em Geophys. J. R. Astr. Soc.} {\bf 13} 529 (1967).


\bibitem{Podlubny}
Podlubny I \emph {Fractional Differential Equations} (Academic
Press, New York, 1999).

\bibitem{Kubo}
Kubo R, {\em Rep. Progr. Phys.} {\bf  29} 255 (1966).


\bibitem{Abramowitz}
Abramowitz M, Stegun I, {\em Handbook of Mathematical Functions} 
(Dover, New York, 1964).

\bibitem{UMB}
Marconi UMB, Puglisi A, Rondoni L, Vulpiani A Fluctuation-dissipation: response theory in statistical physics. {\em Phys. Rep.} {\bf 461} 111 (2008).


\bibitem{Villamaina}
Villamaina D, Puglisi A, Vulpiani A, {\em J. Stat. Mech. L10001} 1 (2008). 


\bibitem{EPL}
Taloni A, Chechkin A, Klafter J, to be published. 


\bibitem{Saxton}
Saxton MJ, {\em Fundamental concepts in biophysics: Volume 1.} Handbook of modern biophysics, ed. T. Jue (Humana Press, New York, 2009).

\bibitem{ripple}
Lee CH, Tsai FC, Wang CC, Lee CH, {\em Phys. Rev. Lett.} {\bf 103}, 238101 (2009).

\bibitem{Champeney}
Champeney DC,   {\em Fourier Transforms and Physical Applications}  (Academic Press, London, 1937).


\bibitem{Olver}
Olver FWJ, {\em Asymptotics and Special Functions} (Academic Press, New York, 1974).


\bibitem{Hardy}
Hardy GH, {\em Divergent Series} (Clarendon Press, Oxford, 1949).




\end{thebibliography}
\end{document}